\documentclass[aps,pra,twocolumn]{revtex4-1}
\usepackage[utf8]{inputenc}
\usepackage{graphicx}
\usepackage{multirow}
\usepackage[colorlinks, citecolor={blue}, urlcolor={blue}, linkcolor={blue}]{hyperref}
\usepackage{amsmath}
\usepackage[dvipsnames]{xcolor}

\begin{document}

\title{The Character of Motional Modes for Entanglement and Sympathetic Cooling \\of Mixed-Species Trapped Ion Chains}

\author{K. Sosnova}
\email{Corresponding Author: ksosnova@umd.edu}
\affiliation{Joint Quantum Institute, Center for Quantum Information and Computer Science, and Department of Physics, University of Maryland, College Park, MD 20742}

\author{A. Carter}
\affiliation{Joint Quantum Institute, Center for Quantum Information and Computer Science, and Department of Physics, University of Maryland, College Park, MD 20742}

\author{C. Monroe}
\affiliation{Joint Quantum Institute, Center for Quantum Information and Computer Science, and Department of Physics, University of Maryland, College Park, MD 20742}

\date{\today}

\begin{abstract}
Modular mixed-species ion-trap networks are a promising framework for scalable quantum information processing, where one species acts as a memory qubit and another as a communication qubit. This architecture requires high-fidelity mixed-species entangling gates to transfer information from communication to memory qubits through their collective motion. We investigate the character of the motional modes of a mixed-species ion chain for entangling operations and also sympathetic cooling. 
We find that the laser power required for high-fidelity entangling gates based on transverse modes is at least an order of magnitude higher than that based on axial modes for widely different masses of the two species. We also find that for even moderate mass differences, the transverse modes are much harder to cool than the axial modes regardless of the ion chain configuration. Therefore, transverse modes conventionally used for operations in single-species ion chains may not be well suited for mixed-species chains with widely different masses.
\end{abstract}

\maketitle

\section{Introduction}
Ion traps are currently among the highest-performing systems for quantum information processing \cite{wineland2008entangled,monroe2013scaling}.  They exhibit nearly perfect idle qubit properties, are highly replicable, and can support high fidelity quantum gate operations~\cite{Ballance:2016, Gaebler2016high}. One of the most promising directions for scaling up such systems is a modular architecture consisting of separate collections of ion crystals with both memory/processing qubits and communication qubits, which provide photonic links between the modules to establish remote entanglement~\cite{Duan2010,Li_2012,Monroe2014}. 

Eliminating crosstalk between communication qubits and memory qubits will likely require the co-trapping of distinct atomic ion species.  
The use of mixed-species ion crystals also allows sympathetic cooling \cite{Larson1986, Morigi2001, Barrett2003, Home2009, Jost2009} of one species (the coolant ions) without affecting the qubit state of the other species (memory ions).  This may be necessary to quench the heating of ion motion \cite{Turchette2000,Brownnutt2015} or remove the excess energy from the shuttling, separation, and recombination of ion strings~\cite{Blakestad2009,Shu2014}. The ability to cool the memory qubits mid-course allows longer computational times and operations with higher fidelity, leading to a wider range of applications for quantum information processing. 

This use of complementary ion species has previously been exploited for high performance atomic clocks, where one species hosts well-isolated levels as the clock that is measured through entanglement and subsequent fluorescence detection of a second species \cite{Schmidt2005}.  In a further similar example, the spectrum of a single molecular ion has been measured through entanglement with a neighboring fluorescing atomic ion \cite{Wolf2016}.

Here, we consider mixed-species ion crystals where one species acts as a memory qubit and the other as a communication qubit, with complementary features.  Once remote entanglement between communication ions is established via the photonic links \cite{Duan2010,Li_2012,Monroe2014}, the information is swapped from a communication ion to one of the memory ions within the same module. The performance of the same procedure in the other node results in entanglement of the two memory qubits in the separate nodes. This entanglement swapping scheme requires two Ising-XX gates~\cite{Molmer1999, Milburn2000, solano1999} with appropriate relative phase control of the two gates~\cite{Hucul2015}. A high-fidelity mixed-species entangling gate is thus an essential ingredient for future quantum networks. Interspecies entangling quantum gates have been performed in various mixed-species ion systems, including $^{9}\textrm{Be}^+$/$^{25}\textrm{Mg}^+$~\cite{Tan2015}, $^{9}\textrm{Be}^+$/$^{40}\textrm{Ca}^+$~\cite{Negnevitsky2018}, $^{40}\textrm{Ca}^+$/$^{88}\textrm{Sr}^+$~\cite{Bruzewicz2019},
$^{171}\textrm{Yb}^+$/$^{138}\textrm{Ba}^+$~\cite{Inlek2017}, and $^{40}\textrm{Ca}^+$/$^{43}\textrm{Ca}^+$~\cite{Ballance2015}.

In our study, $^{171}$Yb$^+$ ions are used for quantum memory and processing because they are insensitive to magnetic field and have long coherence times~\cite{Olmschenk2007, Wang2017}, while 
$^{138}$Ba$^+$ ions are considered as communication qubits since their visible photon-emission lines at 493~nm are more efficient with current fiber-optics and detector technologies~\cite{Auchter2014,Yum17,Araneda2018,Crocker19}.

Quantum entangling gates are mediated by Coulomb collective phonon modes of motion via qubit state-dependent forces. In order to decouple the internal qubit states from the motional states at the end of a gate, it is necessary to know the frequencies and normal modes of motion, and satisfy all of the spin-motion decoupling conditions discussed below [after Eq.~(\ref{eq:alpha})]. To meet these requirements, amplitude~\cite{Zhu2006, ZhuPRL2006, Roos2008, Choi2014,Steane2014}, frequency~\cite{LeungPRL2018,Leung2018, Landsman2019}, or phase~\cite{Green2015, Milne2018, Lu2019} modulation of the driving laser fields, as well as multitone gates~\cite{Shapira2018, Webb2018, Haddadfarshi2016} can be utilized. Each of these methods has been proposed for and implemented in long single-species ion chains.

In the present paper, we discuss the role of axial and transverse normal modes for entangling gates and sympathetic cooling of mixed-species ion chains. First, we perform calculations for the amplitude modulation (AM) and frequency modulation (FM) of the driving laser fields. We optimize the laser amplitude and frequency profiles, respectively, to achieve high fidelity Ising-XX  entangling gates between different species within a long mixed-species ion chain. We compare the suitability of axial and transverse modes for AM and FM pulse-shaping schemes. In the case of transverse modes, the amount of laser power required to satisfy the spin-motion decoupling conditions is very difficult to achieve in a real experiment, because there is a strong transverse mode participation mismatch between different species. The required laser power in the case of axial modes is at least one order of magnitude lower and is routinely achievable in experiments. Therefore, axial modes are preferable for mixed-species gates. The calculations presented in this paper consider $^{171}$Yb$^+$/$^{138}$Ba$^+$ five-ion chains. However, the results of the calculations apply to both shorter and longer mixed-species ion chains with ions of masses that differ by more than 10\%~\cite{Tan2015,Negnevitsky2018,Bruzewicz2019}.

Finally, we study sympathetic cooling in mixed-species chains and discuss how this cooling process depends on normal modes. For species with highly disparate masses, the transverse modes are much harder to cool than axial modes regardless of the configuration of the ions. It is crucial, however, to be able to cool the modes used for entangling gates in a given quantum computing procedure.
We find that in the case of $^{171}$Yb$^+$/$^{138}$Ba$^+$ chains, the mass disparity is significant, and the sympathetic cooling of the transverse modes is inefficient. For $^{171}$Yb$^+$ processing/memory qubits, $^{172}$Yb$^+$ or $^{174}$Yb$^+$ ions would instead be preferred for sympathetic cooling. Note, however, that these ions may not be as suitable for quantum network communication due to spectral overlap and high attenuation of UV light associated with Yb$^+$ in fibers and other photonic components.

\section{Normal mode participation } \label{sec:participation}
We start by considering the motional normal modes in long mixed-species ion chains and their role in mixed-species entangling gates. Consider a long chain of $N$ ions with charge $e$ and different masses $m_j\, (j=1,\dots, N)$, in a linear Paul trap~\cite{James1998, Morigi2001, Home2013}. The dynamics of the system are described by the Lagrangian
\begin{equation}
\mathcal{L} =  \sum_{i=1}^N\dfrac{m_i\dot{\boldsymbol{r}}_i^2}{2}-U,
\end{equation}
where $U$ is the potential energy:
\begin{equation}
U =  \sum_{i=1}^N\Phi(\boldsymbol{r}_i, m_i) +\dfrac{1}{2}\sum_{\substack{i,j=1 \\ i\neq j}}^N
\dfrac{e^2}{4\pi\epsilon_0\vert \boldsymbol{r}_i - \boldsymbol{r}_j\vert}.
\end{equation}
Here $\Phi(\boldsymbol{r}_i, m_i)$ is the potential energy of an ion $i$ with the mass $m_i$ at a position $\boldsymbol{r}_i$ in the harmonic potential of the trap electrodes. 

The equilibrium position of ion $i$ along the trap axis, $z_i^{(0)}$, is determined by setting $\partial U/\partial z_i = 0$. By symmetry, $x_i^{(0)}\nobreak=\nobreak0$ and $y_i^{(0)}=0$.
The standard Taylor expansion of the Lagrangian around the equilibrium positions yields
\begin{eqnarray}
    \mathcal{L} \approx  \dfrac{1}{2}\left(\sum_{i=1}^N\right. m_i&\dot{q_i}^2& -\left.\sum_{i,j=1}^N V_{ij}q_iq_j \right),\\ \nonumber
    V_{ij} &=& \left.\dfrac{\partial^2 U}{\partial q_i\partial q_j}\right\vert_0.
\end{eqnarray}
For axial modes, $q_i = z_i-z_i^{(0)}$, while for transverse modes, $q_i = x_i$ or $y_i$. The Lagrange equations for the normal modes of motion are then given by:
\begin{equation} \label{eq:eigen}
    \sum_{i=1}^N V_{ij}b_{im} = \lambda_m m_i b_{im},
\end{equation}
where $\lambda_m = \omega_m^2$ is the eigenvalue of the $m^{\textrm{th}}$ mode, with $\omega_m$ the mode frequency and $b_{im}$ the normal mode transformation matrix element between ion $i$ and mode $m$ with $\sum_i b_{im}b_{in} = \delta_{nm}$ and $\sum_m b_{im}b_{jm} = \delta_{ij}$. Each normal mode represents an individual harmonic oscillator that can be quantized. We introduce the creation and annihilation operators, $\hat{a}_m^{\dagger}$ and $\hat{a}_m$, for the mode $m$, and the original set of coordinates assumes the following standard quantized form:
\begin{equation} \label{eq:xj}
    \hat{x}_i = \sum_{m=1}^N b_{im} \sqrt{\dfrac{\hbar}{2m_i\omega_m}}\left( \hat{a}_m+\hat{a}_m^{\dagger} \right).
\end{equation}

We start our discussion of multi-ion entangling quantum gates by introducing a generic laser-ion interaction Hamiltonian:
\begin{equation} \label{eq:Hint}
    H_I = \hbar\sum_i \left( \Omega_i e^{i\mu t}e^{i(\Delta k x_i+\Delta \phi_i)}\hat{\sigma}_i^+ + {\rm H.c.} \right),
\end{equation}
where $\Omega_i$ is the resonant Rabi frequency of the $i^{\textrm{th}}$ ion. Assuming a two-photon Raman coupling \cite{wineland2008entangled}, we use two laser beams with wave-vector difference $\Delta k$, frequency difference $\mu$, and phase difference $\Delta \phi_i$. After substituting Eq.~(\ref{eq:xj}) into Eq.~(\ref{eq:Hint}), one can separate out the standard Lamb-Dicke parameters $\eta_{im}$ in the exponents:
\begin{equation} \label{eq:eta}
    \eta_{im} = |\Delta k|b_{im} \sqrt{\dfrac{\hbar}{2m_i\omega_m}}.
\end{equation}
We create the M{\o}lmer-S{\o}rensen or Ising interaction~\cite{Molmer1999, Milburn2000, solano1999} by simultaneously driving off-resonant red and blue sideband transitions on each of the qubits.
The corresponding evolution operator has the following form~\cite{Zhu2003}:
\begin{eqnarray}
    U(\tau) = \textrm{exp}\left\{ \sum_{i,m}\left[ \alpha_{im}(\tau)\hat{a}_m^{\dagger} - \alpha^*_{im}(\tau)\hat{a}_m \right]\hat{\sigma}_i^x  \right.\\ \nonumber \left.+\, i\sum_{i,j}\chi_{ij}\hat{\sigma}_i^x\hat{\sigma}_j^x \right\},
\end{eqnarray}
where
\begin{eqnarray} \label{eq:alpha}
 \alpha_{im}(\tau) &=& -\int_0^{\tau} \eta_{im}\Omega_i(t)\textrm{exp}(i\omega_m t)dt, \\ \nonumber
  \chi_{ij}(\tau) &=& \sum_m \eta_{im} \eta_{jm} \int_0^{\tau}dt_1\int_0^{t_1}dt_2\; \textrm{sin}\left[\omega_m (t_1-t_2)\right]\\ \nonumber
  &\times& \left[ \Omega_i(t_1)\Omega_j(t_2)+\Omega_j(t_1)\Omega_i(t_2) \right].
\end{eqnarray}
To be able to drive a fully-entangling gate between ions $i$ and $j$ in time $\tau$, we require $\chi_{ij}(\tau) = \pi/4$. In order to decouple the motional and the spin degrees of freedom by the end of the gate evolution, we also require $\alpha_{im}(\tau) = 0$ for each ion $i$ and mode $m$. Therefore, there are $2N+1$ constraints for a perfect gate operation: one phase condition and $2N$ spin-motion decoupling conditions (counting independently the real and imaginary components that correspond to coordinates and momenta, respectively). In this case, the evolution operator $U(\tau)$ reduces to $U_{ij} = \textrm{exp}(i\pi\hat{\sigma}_i^x\hat{\sigma}_j^x/4)$. As mentioned previously, a number of methods have been introduced to fulfill these requirements. Among them are pulse-shaping techniques -- using amplitude~\cite{Zhu2006, ZhuPRL2006, Roos2008, Choi2014,Steane2014}, frequency ~\cite{LeungPRL2018,Leung2018, Landsman2019}, or phase~\cite{Green2015, Milne2018, Lu2019} modulation of the driving laser fields, -- as well as multitone gates~\cite{Shapira2018, Webb2018, Haddadfarshi2016}.

The expressions for $\alpha_{im}(\tau)$ and $\chi_{ij}(\tau)$ both depend on the normal modes from Eq.~(\ref{eq:eigen}). In the present manuscript, we consider multi-species ion chains with 5 and 13 ions for visual clarity, but we tested all our findings and conclusions in longer ion chains of up to 50 ions. 

\begin{figure}
 \includegraphics[width=0.82\linewidth]{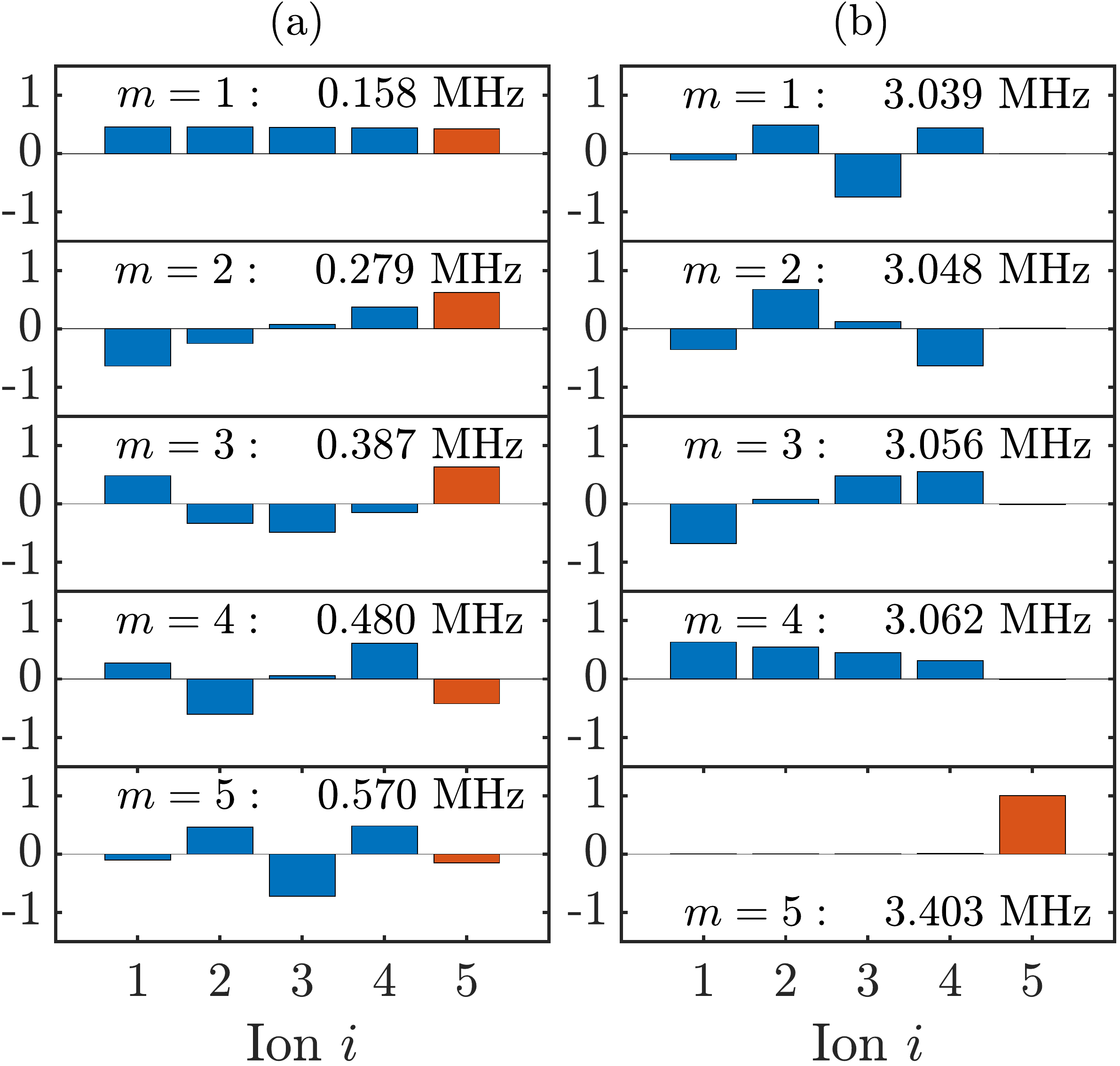}
 \caption{\label{fig_modes} (a) Axial normal modes and their frequencies for a five-ion chain composed of four $^{171}$Yb$^+$ ions (blue) and one $^{138}$Ba$^+$ ion (red) located at the edge of the ion chain. (b) The same for the transverse normal modes.}
\end{figure}
In Fig.~\ref{fig_modes}, we show the frequencies $\omega_m$ and the normal modes $b_{im}$ for a five-ion chain with four $^{171}$Yb$^+$ ions and one $^{138}$Ba$^+$ ion located at the edge of the chain. We find this configuration the most experimentally convenient for individual addressing and reordering purposes. Moreover, we must avoid a configuration with the $^{138}$Ba$^+$ ion in the center of the chain, since, due to the reflection symmetry, there are modes in which the $^{138}$Ba$^+$ ion does not participate at all.
The experimentally accessible trapping parameters used in the calculations are the transverse trapping frequency $\omega_x/2\pi = 3.06$~MHz~\cite{Choi2014} and the axial trapping frequency $\omega_z/2\pi = 0.16$~MHz. The axial frequency is chosen to be relatively low in order to maintain linear ion chains for up to 50 ions.

\begin{table}[t]
\caption{\label{table:AM5} Maximum Rabi frequencies $\Omega_{\rm max}$ and average fidelities $F$ of five-segment AM entangling gates between different $^{171}$Yb$^+/^{138}$Ba$^+$ pairs in a five-ion chain based on axial and transverse modes. The single $^{138}$Ba$^+$ ion is located at position 5.}
\begin{tabular}{ |p{1.5cm}||p{1.6cm}|p{1.5cm}|p{1.6cm}| p{1.5cm}|  }
 \hline
 &\multicolumn{2}{|c|}{Axial gates} & \multicolumn{2}{|c|}{Transverse gates}\\
 \hline
 \centering Ion pair& \centering $\frac{\Omega_{\rm max}}{2\pi}$(kHz) & \centering $F$ (\%)  & \centering $\frac{\Omega_{\rm max}}{2\pi}$(kHz) & $F$ (\%)\\
 \hline
  \centering $(1, 5)$ & \centering 11 & \centering 99.86 & \centering 400 & 81.03 \\
  \centering $(2, 5)$ & \centering 11 & \centering 99.86 & \centering 310 & 85.86 \\
  \centering $(3, 5)$ & \centering 12 & \centering 99.82 & \centering 175 & 95.65 \\
  \centering $(4, 5)$ & \centering  9 & \centering 99.81 & \centering  70  & 99.22 \\
 \hline
\end{tabular}
\end{table}

The axial normal modes and frequencies for a chain of four $^{171}$Yb$^+$ ions and one $^{138}$Ba$^+$ ion [shown in Fig.~\ref{fig_modes}(a)] do not differ much from those in a pure five $\,^{171}$Yb$^+$ chain. This indicates that entangling gates will work as efficiently as in a chain of five $^{171}$Yb$^+$ ions. On the other hand, the transverse normal modes and frequencies differ significantly from those in the pure five $\,^{171}$Yb$^+$ chain. A drastic mismatch can be seen, for example, in the center-of-mass (COM) mode -- the mode with the highest frequency in the bottom panel in Fig.~\ref{fig_modes}(b). Given a moderate mass disparity, we expect the amplitudes of motion for all ions in the COM mode to be close to each other, as in the axial case. For the transverse modes, however, the $^{138}$Ba$^+$ ion motion decouples from that of the $^{171}$Yb$^+$ ions. Furthermore, the greater the ion mass disparity, the larger the calculated mismatch. Also note that the highest frequency $\omega_5/2\pi = 3.403$~MHz is distant and isolated from the rest of the frequencies with a gap of 340~kHz, while the average frequency difference is about 8~kHz. This fact also is relevant to the discussion of sympathetic cooling in Sec.~\ref{sec:sympath}.

The transverse modes present difficulties associated with the mode mismatch. However, they are preferable for quantum entangling gate operations because their mode frequencies are higher, which allows better cooling, less susceptibility to heating, and faster gates.  The coupling to transverse modes also allows individual addressing~\cite{Debnath2016}. We therefore focus in this manuscript on various techniques that might allow us to perform fast transverse entangling mixed-species gates.

\section{AM gates}\label{sec:AM}
First, we consider amplitude modulation (AM) of the driving field~\cite{Zhu2006, ZhuPRL2006, Roos2008,Choi2014,Steane2014} to satisfy the $2N+1$ conditions discussed in Sec.~\ref{sec:participation}. In particular, the Rabi frequency $\Omega(t)$ is modulated in time as a piecewise-constant segmented pulse defined as:
\begin{equation} \label{eq:AM_shaping}
  \Omega(t) =
    \begin{cases}
     \Omega_1, \;\;\;\;\;\;\;\; 0\leq t \leq \tau/P \\
      \Omega_2, \;\;\; \tau/P\leq t \leq 2\tau/P\\
    \;\;  \vdots \\
      \Omega_P, \;\;\; (P-1)\tau/P\leq t \leq \tau
    \end{cases},
\end{equation}
where $\tau$ is the total gate time.

If $\Omega(t)$ has $2N+1$ segments, the existence of the solution for $2N+1$ constraints is guaranteed, and the problem is reduced to a system of linear equations. However, for long single-species chains, one can use fewer than $N$ segments and still achieve high fidelity entangling gates, since most of the motional modes have relatively low populations and needn't decouple perfectly~\cite{Landsman2019}. As discussed in Sec.~\ref{sec:participation}, the axial normal modes in the ($4^{171}$Yb$^+-^{138}$Ba$^+$) chain are similar to those in the pure five $\,^{171}$Yb$^+$ chain, and all the findings and techniques used for single-species chains are expected to apply.

\begin{figure}
 \includegraphics[width=0.7\linewidth]{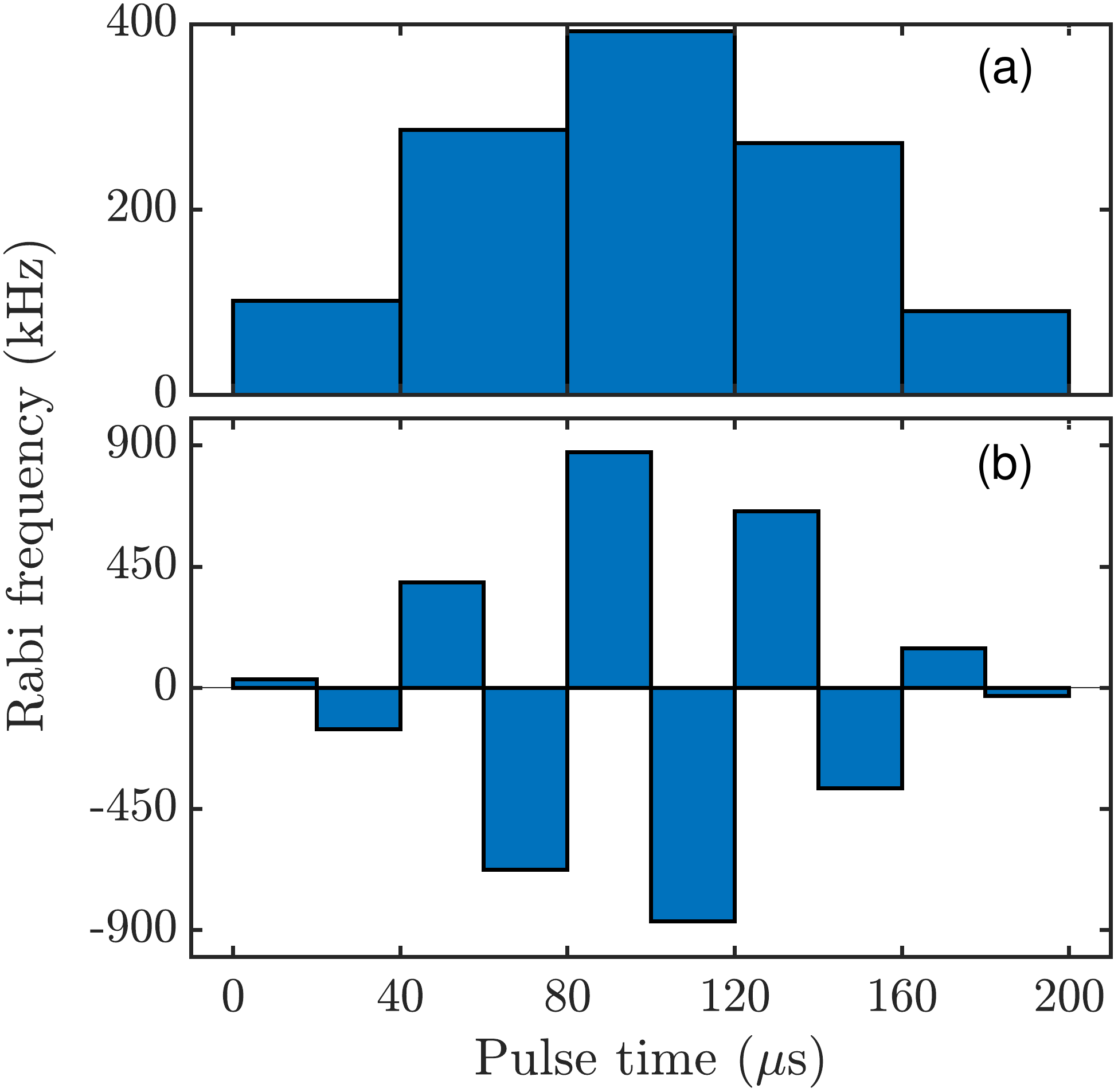}
 \caption{\label{fig_5-10_seg} Amplitude-modulated pulses of the driving field for an AM entangling gate between $^{171}$Yb$^+$ and $^{138}$Ba$^+$ ions in a five-ion chain for the ion pair $(1, 5)$. (a) Five-segment pulse with maximum Rabi frequency $\Omega_{\rm max}/2\pi = 400$~kHz; (b) ten-segment pulse with the maximum Rabi frequency $\Omega_{\rm max}/2\pi\nobreak=\nobreak900$~kHz. In both cases, the pulse duration is $\tau = 200~\mu$s.}
\end{figure}

In Table~\ref{table:AM5}, we show the calculations for a five-segment AM pulse applied to a five-ion (four $^{171}$Yb$^+$ and one $^{138}$Ba$^+$) chain with a pulse duration $\tau\nobreak=\nobreak200\mu$s. The $^{138}$Ba$^+$ ion is located at the edge of the chain -- see Fig.~\ref{fig_modes}. As shown in the table, for axial gates, relatively low Rabi frequencies -- of the order of 10~kHz -- are sufficient to achieve high-fidelity entangling gates between any pair of qubits. On the contrary, for transverse gates, the required Rabi frequencies are much higher, while the fidelities are much lower. This drastic difference is due to the large mismatch between $^{138}$Ba$^+$ and $^{171}$Yb$^+$ participation in the transverse modes, as shown in Fig.~\ref{fig_modes}. The ion pair $(1, 5)$ has the worst amplitude mismatch, which leads to the highest $\Omega_{\rm max}$ and a relatively low fidelity. Obviously, five segments of the AM pulse are not enough to perform a high-fidelity transverse gate in the pair $(1, 5)$. By increasing the segment number to $2N=10$, we obtain an average gate fidelity $F=99.996\%$. However, in this case, an even higher Rabi frequency is required [see Fig.~\ref{fig_5-10_seg}(b)]. To be able to perform high-fidelity entangling transverse gates between $^{171}$Yb$^+$ and $^{138}$Ba$^+$, we require a Rabi frequency about 80 times higher than for axial gates between the same pair of ions, and the corresponding intensities are not feasible in real experiments. We therefore hereafter focus on the most difficult case -- entangling gates between the ion pair $(1, 5)$ having the largest transverse mode mismatch.

\begin{figure}
 \includegraphics[width=\linewidth]{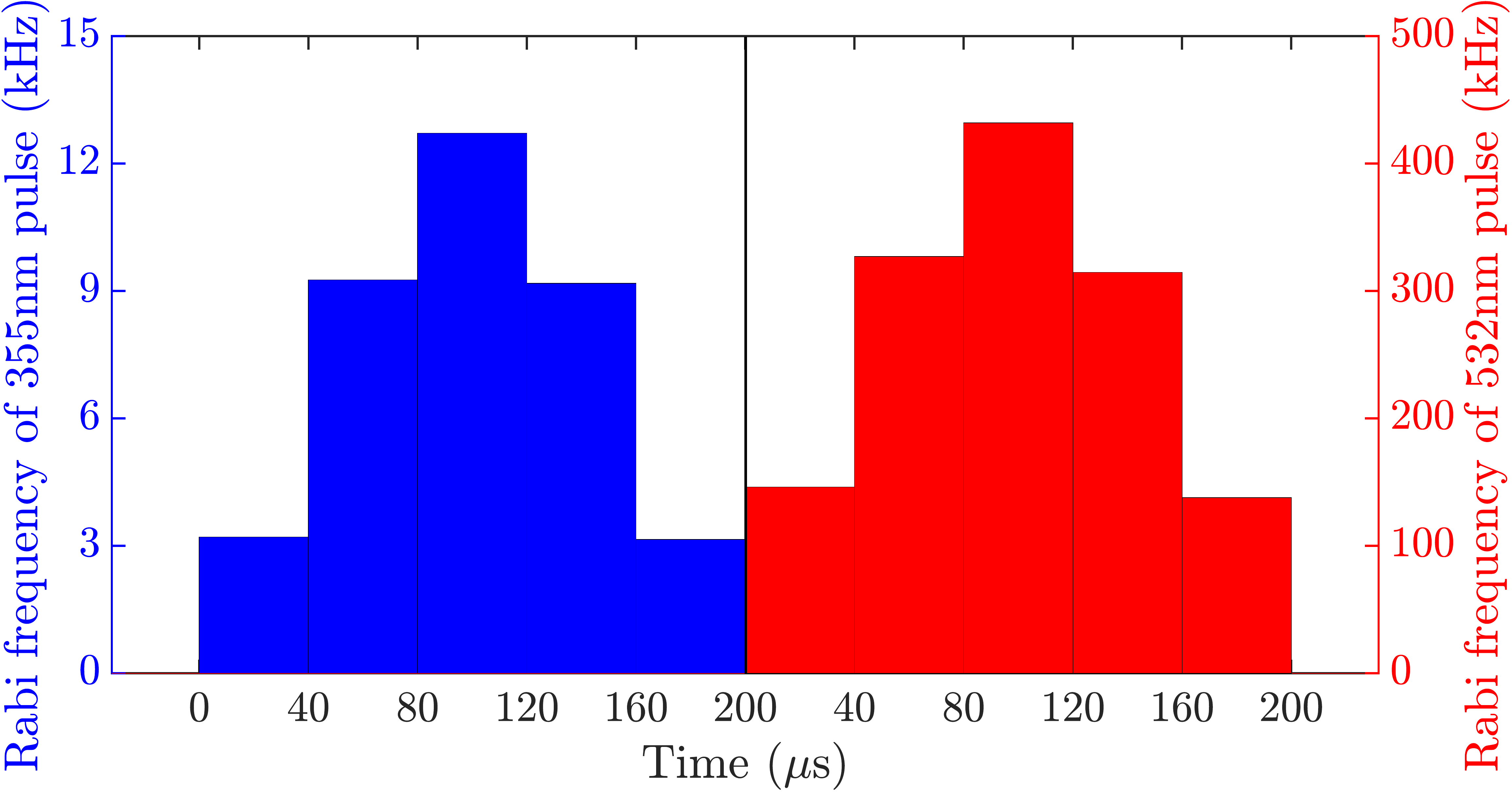}
 \caption{\label{fig_diffRabi} Dual five-segment amplitude modulation of the driving fields for an entangling gate between $^{171}$Yb$^+$ and $^{138}$Ba$^+$ ions in a five-ion chain, ion pair $(1, 5)$. 355~nm AM pulse shape and the corresponding vertical axis on the left side are shown in blue; 532~nm pulse and the corresponding vertical axis on the right side are in red. The pulse duration $\tau = 200~\mu$s.}
\end{figure}

In our experimental setup, we perform rotations in $^{171}$Yb$^+$ using 355~nm light from the Spectra Physics Vanguard pulsed laser~\cite{Hayes2010}. This light is the third harmonic of a 1064~nm Nd:YVO4 source. Conveniently, the second harmonic at 532~nm from the same laser can be used to drive rotations in $^{138}$Ba$^+$~\cite{Inlek2017}. Previously, we assumed equal intensities for each driving field, and arrived at extremely high laser intensities required to drive the entangling transverse gates between $^{171}$Yb$^+$ and $^{138}$Ba$^+$ ions. However, we can instead apply 355~nm and 532~nm Raman beams with different powers, and thereby generalize the conventional AM pulse-shaping technique. We refer to this approach as dual AM pulse shaping, and modify Eq.~(\ref{eq:AM_shaping}) to allow different values of $\Omega^{355}_j$ and $\Omega^{532}_j$, $j=1,\ldots,P$ and perform optimization of these $2P$ independent parameters to find $\Omega^{355}(t)$ and $\Omega^{532}(t)$. 

We simulate the five-segment entangling transverse gate between $^{171}$Yb$^+$ and $^{138}$Ba$^+$ ions in the five-ion chain, but due to independent intensities, still have $2N$ parameters to vary. The optimization results are presented in Fig.~\ref{fig_diffRabi}. In the case of independent Rabi frequencies, since we have twice as many degrees of freedom, the fidelity of the quantum operation $F = 95.22\%$ is significantly higher than in the conventional five-segment AM pulse-shaping technique ($81.03\%$, see Table~\ref{table:AM5}). The optimization procedure finds the gate frequency $\mu \approx \omega_4 + 2\pi\times8$~kHz, which is close to the lower four modes and far from the highest, isolated mode [see the mode spectrum in Fig.~\ref{fig_modes}(b)]. Also, since the transverse modes have a strong participation mismatch for the pair $(1, 5)$ at the focus of the calculations, the Rabi frequencies of the 355~nm and the 532~nm pulses are widely disparate with the 532~nm Rabi frequency difficult to achieve in the laboratory: $\Omega^{355}_{\rm max}/2\pi = 13$~kHz and $\Omega^{532}_{\rm max}/2\pi = 420$~kHz. For comparison, we also perform calculations for the seven-segment dual AM pulse. We improved the gate fidelity to $F = 99.53\%$ with similar Rabi frequencies, $\Omega^{355}_{\rm max}/2\pi = 20$~kHz and $\Omega^{532}_{\rm max}/2\pi=440$~kHz, while the conventional seven-segment AM model gives us $F = 96.5\%$ with $\Omega_{\rm max}/2\pi = 540$~kHz. By increasing the number of segments in the dual AM pulse further, we achieve higher gate fidelities, but not lower Rabi frequencies.

\section{AM-FM gates}\label{sec:AMFM}
As discussed in Sec.~\ref{sec:AM}, the AM transverse gates require driving fields with relatively high Rabi frequencies. In this section, we first consider a different way of satisfying the $2N+1$ conditions (the $\pi/4$ phase condition and the spin-motion decoupling conditions): frequency modulation (FM) of the driving fields~\cite{LeungPRL2018,Leung2018, Landsman2019}. In this case, we allow the frequency of the driving field to vary in time, while the resonant Rabi frequency $\Omega(t)=\Omega$ from Eq. \ref{eq:alpha} is constant in time. In this case, $\alpha_{im}(\tau)$ and $\chi_{ij}(\tau)$ take the following form~\cite{LeungPRL2018}:

\begin{eqnarray}
 \alpha_{im}(\tau) &=& -\eta_{im}\Omega\int_0^{\tau} e^{i\theta_m(t)}dt, \\ \nonumber
  \chi_{ij}(\tau) &=& \Omega^2\!\sum_{m=1}^N\!\eta_{im} \eta_{jm}\!\!\int_0^{\tau}\!\!\!dt_1\!\!\int_0^{t_1}\!\!\!dt_2\,\textrm{sin}\left[\theta_m(t_1)-\theta_m(t_2)\right],\\ \nonumber
 \theta_m(t) &=& \int_0^t \delta_m(t')dt',
\end{eqnarray}
where $\delta_m(t)$ is the detuning of the driving field relative to the mode $m$, and we assume equal Rabi frequencies on all ions. First, we try to satisfy $2N$ conditions $\alpha_{im}(\tau) = 0$ by optimizing a cost function defined by the sum of squares of time-averaged displacements of the phase-space trajectories of the modes (see Ref.~\cite{LeungPRL2018}). Once the frequency profile is found, we choose the Rabi frequency $\Omega_{\rm max}$ to satisfy the remaining entanglement condition $\chi_{ij}(\tau) =\pi/4$ for a given pair of qubits.

\begin{figure}
 \includegraphics[width=\linewidth]{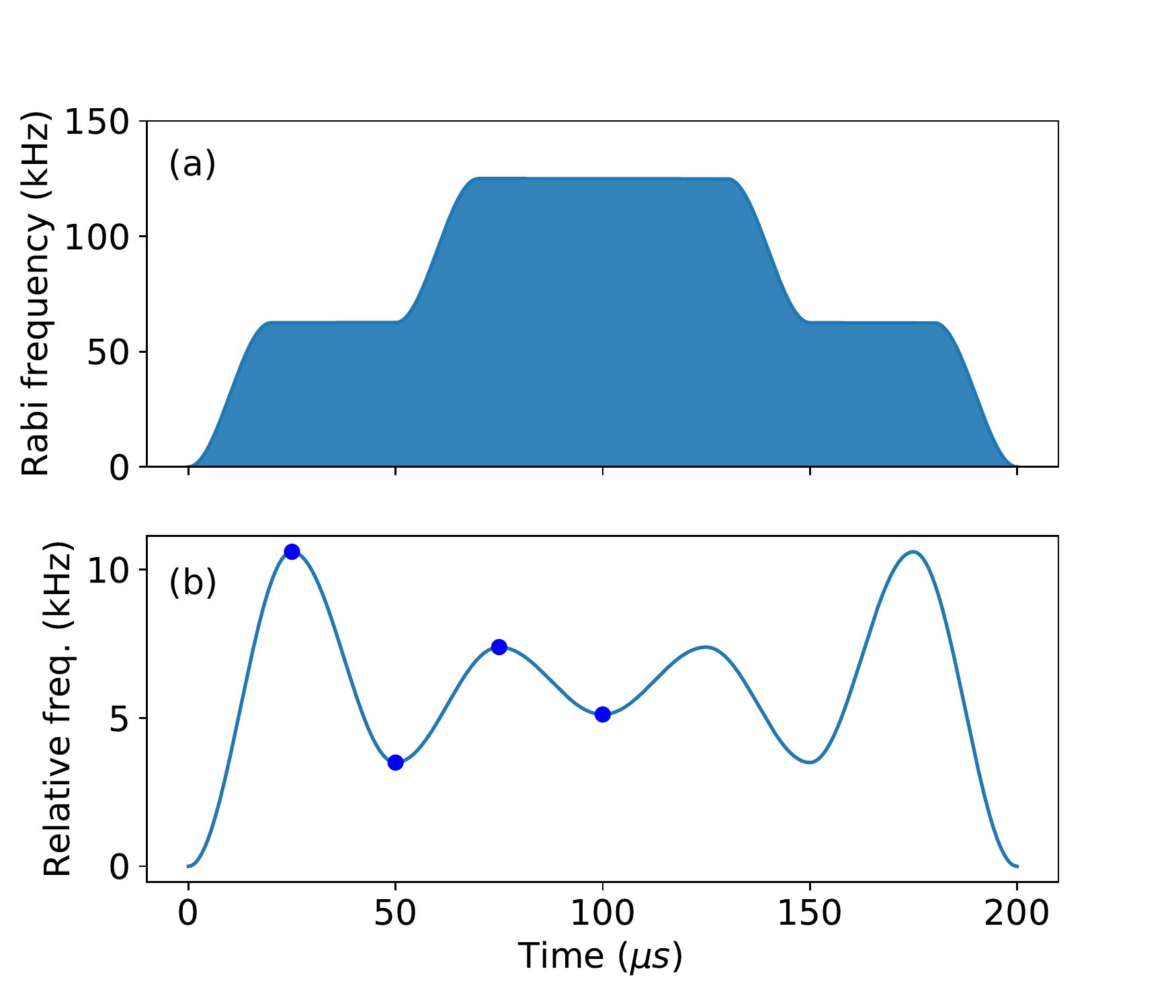}
 \caption{\label{fig_FM} Amplitude and frequency modulation of the driving fields for an AM-FM entangling gate between $^{171}$Yb$^+$ and $^{138}$Ba$^+$ ions in a five-ion chain -- ion pair $(1, 5)$. (a) Fixed amplitude modulation consists of three plateaus connected with cosine ramps. (b) Optimized frequency modulation has a set of turning points (blue dots) connected via cosine curves. Note that the pulse is set to be symmetric in time. The gate duration $\tau = 200~\mu$s.}
\end{figure}

Following Ref.~\cite{Leung2018}, we choose the FM pulse shape to be symmetric in time and combine this frequency modulation optimization with a fixed amplitude modulation $\Omega(t)$ of the shape presented in Fig.~\ref{fig_FM}(a) (three plateaus connected by smooth cosine ramps). The resulting scheme is referred to as an AM-FM gate.

Not surprisingly, similarly to the case in Sec.~\ref{sec:AM}, high-fidelity axial gates between any pair of ions in the mixed-species chain require low Rabi frequencies of the order of 10~kHz. The transverse entangling gates, however, still require much higher Rabi frequencies of the driving fields due to the strong amplitude mismatch between the $^{171}$Yb$^+$ and $^{138}$Ba$^+$ ions, despite the more sophisticated driving scheme.

In Fig.~\ref{fig_FM}, we show the frequency and amplitude modulation of the driving laser fields for the AM-FM entangling operation between $^{171}$Yb$^+$ and $^{138}$Ba$^+$ ions for the ion pair $(1, 5)$. As shown in panel (b), we first try a pulse with four distinct turning points in the frequency domain [the intensity-domain shape is fixed to that shown in panel (a)]. For this example, as an initial guess, we choose the reference frequency $\mu = \omega_4 + 2\pi\times2.5$~kHz that represents the 0 level in panel (b). The fidelity of this entangling gate is $F = 99.76\%$, and the maximum Rabi frequency is $\Omega_{\rm max}/2\pi = 125$~kHz [see Fig.~\ref{fig_FM}(a)]. By increasing the number of turning points in the frequency domain, we achieved a much higher gate fidelity, but it did not allow us to lower the required Rabi frequencies. Although the required Rabi frequency in the AM-FM gate is much lower than that for the AM gate, it is still over one order of magnitude higher than the Rabi frequency necessary for the axial entangling gates. However, while it remains an experimental challenge to achieve the required powers, this approach to transverse gates is much more promising than pure AM or FM techniques.

The calculations presented in this paper are performed for $^{171}$Yb$^+/^{138}$Ba$^+$ five-ion chains. However, our findings are readily generalized to shorter and longer mixed-species ion chains (we checked lengths between 2 and 50 ions) having highly disparate masses. Moreover, the calculations for even longer ion chains can be performed with the appropriate trapping frequencies, so that the effective potential is sufficiently shallow so that the ions remain in a line.
Ion pairs with highly disparate masses include $^{9}$Be$^+/^{25}$Mg$^+$~\cite{Tan2015}, $^{9}$Be$^+/^{40}$Ca$^+$~\cite{Negnevitsky2018}, and $^{40}$Ca$^+/^{88}$Sr$^+$~\cite{Bruzewicz2019}. 
In fact, mixed-species entangling gates in two-ion crystals have been demonstrated experimentally~\cite{Tan2015, Inlek2017, Negnevitsky2018,Bruzewicz2019}, but exclusively with axial entangling gates due to the strong transverse mode mismatch. Axial gates were even used for the similar masses of a $^{40}$Ca$^+/^{43}$Ca$^+$ chain~\cite{Ballance2015}.

\section{Sympathetic cooling} \label{sec:sympath}
In Secs.~\ref{sec:AM}--\ref{sec:AMFM}, we discussed the role of the normal modes in the mixed-species entangling operations in long ion chains. These operations are a necessary element for modular quantum networks. Another important application of mixed-species ion chains is sympathetic cooling~\cite{Larson1986, Morigi2001, Barrett2003, Home2009, Jost2009}. We can constantly cool certain ions in the chain (``coolant'' ions) -- $^{138}$Ba$^+$ in this case -- while continuously performing quantum computations with the processing ions -- $^{171}$Yb$^+$ in this case~\cite{Wang2017}.

One possible characterization of cooling is to consider the root-mean-square position fluctuation $\delta q_i = \sqrt{\langle q_i^2 \rangle}$ of the ions from equilibrium~\cite{Lin2016} with $q_i = x_i$ for the transverse modes and $q_i\nobreak=\nobreak z_i\nobreak-\nobreak z_i^{(0)}$ for the axial modes. Using Eq.~(\ref{eq:xj}), we obtain:
\begin{equation}\label{eq:flucts}
    \langle q_i^2\rangle = \sum_m \dfrac{\hbar}{2m_i\omega_m}b_{im}^2\left( \bar{n}_m+\dfrac{1}{2} \right),
\end{equation}
where $\bar{n}_m$ is average (thermal) phonon number in the $m^{\textrm{th}}$ mode.
Since $\delta q_i$ depends critically on the mode structure, we expect the cooling rates for the transverse and axial mixed-species modes to differ dramatically.

A few coolant ion configurations have been proposed for long ion chains~\cite{Lin2016}, including edge cooling (the coolant ions are located at the edges of the chain), and periodic-node cooling (the coolant ions are positioned periodically in the chain). We explored these approaches using the metrics in Eq. (\ref{eq:flucts}) and, as expected, find that the axial modes are easy to cool even with a small number of the coolant ions, while it is hard to cool the transverse modes, especially the higher-frequency ones, due to the strong amplitude mismatch between the different species.

\begin{figure}
 \includegraphics[width=\linewidth]{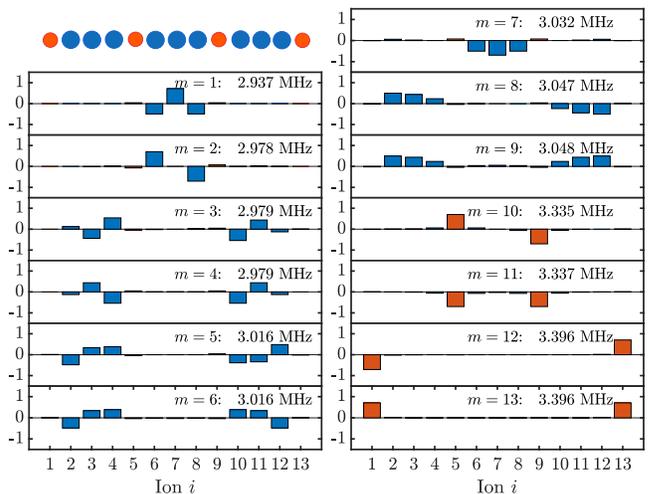}
 \caption{\label{fig_13} Frequencies $\omega_m/2\pi$ and transverse normal mode eigenvectors $b_{im}$ for mode $m$ of ion $i$ in a 13-ion-long chain - nine $^{171}$Yb$^+$ ions (blue) and four $^{138}$Ba$^+$ ion (red) placed periodically in the ion chain. The mode with the highest frequency is the center-of-mass (COM) mode.}
\end{figure}

To get an intuitive understanding of the difficulties in cooling the transverse modes, we show all the transverse modes for a 13-ion chain in Fig.~\ref{fig_13}. In this example, we follow Ref.~\cite{Lin2016}, and place four coolant $^{138}$Ba$^+$ ions periodically in the chain (see Fig.~\ref{fig_13}, top-left panel). The mode with the highest frequency (bottom-right panel) is the COM mode. Only two edge $^{138}$Ba$^+$ ions are moving in this case, while the motion of all $^{171}$Yb$^+$ ions is virtually absent and is thus completely decoupled from the motion of the $^{138}$Ba$^+$ ions. The next three modes have similar behavior. At the same time, the remaining modes have virtually no $^{138}$Ba$^+$ motion. Note that there are four higher frequencies distant from the other nine frequencies with a gap of 350~kHz, while the average frequency difference is about 15~kHz. We have already discussed a similar feature in Sec.~\ref{sec:participation}, where in the ion chain there was only one $^{138}$Ba$^+$ ion, and the highest frequency was isolated from the rest of the spectrum. The number of the isolated spectral lines here is equal to number of the $^{138}$Ba$^+$ ions in the chain as well, and these frequencies correspond to the modes with the most pronounced $^{138}$Ba$^+$ motion.

Due to the high mass ratio between $^{171}$Yb$^+$ and $^{138}$Ba$^+$, we observe a strong amplitude mismatch in the transverse modes, leading to inefficient sympathetic cooling of the transverse modes. These results can be easily generalized to any long ion chain with mixed species. In our case, for the processing $^{171}$Yb$^+$ qubits, suitable cooling ions would be $^{172}$Yb$^+$ or $^{174}$Yb$^+$.

\section{Summary}
Mixed-species chains of atomic ions may be crucial for scaling trapped ion quantum computers and communication networks.  Not only are two species needed for photonic networking between modules, but sympathetic cooling between disparate atomic ions may also be required to quench heating and remove energy from shuttling operations. 

We investigated the role of normal modes in entangling operations and sympathetic cooling in mixed-species ion chains. First, we performed calculations on AM pulse shaping to optimize the fidelity of M{\o}lmer-S{\o}rensen entangling gates between different species based on both axial and transverse modes. 
Due to the strong mismatch in transverse modes between the amplitudes of motion of atomic species having widely disparate masses, the amount of laser power required to satisfy all of the spin-motion decoupling conditions is difficult to achieve in a real experiment. Then we performed calculations for a more advanced, combined AM-FM pulse-shaping approach that allowed considerably smaller Rabi frequencies and experimentally feasible laser powers. However, the laser power required for high-fidelity transverse entangling gates is still at least an order of magnitude higher than that for the axial gates for the case of $^{171}$Yb$^+$ and $^{138}$Ba$^+$. We also find that for mixed species of disparate masses, transverse modes are much harder to cool than axial modes regardless of the configuration of the ions.
We thus conclude that for interspecies gates it will likely be necessary to utilize the axial modes for any operations that involve the motion of the ions.

\section{Acknowledgements}
We thank P. H. Leung, N. Linke, M. Cetina, and L. Egan for helpful discussions. This work was supported by the ARO with funds from the IARPA LogiQ program and the MURI on Modular Quantum Systems, the AFOSR project on Quantum Networks, the ARL Center for Distributed Quantum Information, and the National Science Foundation Physics Frontier Center at JQI.


%

\end{document}